\documentclass[final]{svjour2}
\usepackage{graphicx}
\usepackage{rotating}
\usepackage{amssymb}
\usepackage{mathptmx}
\usepackage{amsmath,amsfonts}
\usepackage[numbers]{natbib}
\makeatletter
\journalname{Journal of Low Temperature Physics}

\def\AA{\mathring{\text{A}}}

\let\a=\alpha \let\b=\beta  \let\d=\delta
   \let\k=\kappa
\let\l=\lambda \let\m=\mu \let\n=\nu  
\let\s=\sigma \let\t=\tau  
   
\let\D=\Delta \let\L=\Lambda  
\let\Si=\Sigma   
\let\ee=\epsilon \let\r=\rho  
\let\om=\omega
\def\ie{{i.e. }}\def\eg{{e.g. }}

\def\DD{{\cal D}}

\def\to{\rightarrow} \def\la{\langle} \def\ra{\rangle}

\newcommand{\beq}{\begin{equation}} \newcommand{\eeq}{\end{equation}}
 \newcommand{\wt}{\widetilde}
\newcommand{\Tr}{\text{Tr}}

\begin{document}

\title{
A tentative replica theory of glassy Helium 4
} 

\author{G.Biroli$^1$ \and F.Zamponi$^2$}
\institute{1: Institut de Physique Th\'eorique (IPhT),
CEA, and CNRS URA 2306, F-91191 Gif-sur-Yvette, France \\
2: Laboratoire de Physique Th\'eorique, UMR 8549, CNRS and Ecole Normale Sup\'erieure, 
24 Rue Lhomond, 75231 Paris Cedex 05, France
}

\date{\today}

\maketitle


\begin{abstract}

We develop a quantum replica method for interacting particle systems and 
use it to estimate the location of the glass transition line in Helium 4. 
Although we do not fully succeed in taking into account all quantum effects,
we make a thorough semiclassical analysis.
We confirm previous suggestions that quantum fluctuations promote the formation
of the glass and give a quantitative estimate of this effect at high density.
Finally, we discuss the difficulties that are met when one tries to extend the
calculation to the region of low densities and low temperatures, where quantum effects
are strong and the semiclassical expansion breaks down.

\end{abstract}

\section{Introduction}

A series of recent experiments on solid He$^4$, initiated by Kim and Chan~\cite{Chan1,Chan2}
and then performed by several groups~\cite{balibarreview},
raised, among many others, the important question of whether disorder and fast quenches
can induce the formation of a quantum glass phase in Helium 4 and to what extent this 
new phase is related to the super-solid behavior observed experimentally. 
Following early numerical and theoretical analyses, that showed that perfect crystals do not support 
supersolidity~\cite{dis1,dis2}, and a numerical analysis that proposed the presence of a superglass
phase at intermediate densities~\cite{BPS1},
Ritner and Reppy~\cite{reppy1,reppy2}
showed that fast quenches produce disordered samples with
a change in the moment of inertia that corresponds to an
extremely high superfluid fraction, of the order of $20\%$.
Since then, whether this behavior can be attributed to the presence of a quantum
(super-)glass phase has been the subject of some 
debate~\cite{norho_s,Ukraine07,Balatsky1,Hunt09,Saunders09,Balatsky2,BCFZ11}.
However, before addressing the difficult question of whether supersolidity is
possible, present and possibly enhanced in quantum glasses, it would be desirable to have a full
understanding of the nature of the glass transition in presence of strong 
quantum fluctuations, a subject which is instead very poorly understood.

Indeed, several authors attempted to build a
theory of the quantum glass transition~\cite{MF1,MF2,MF3,MF4,MF5,BCZ08}, but none
has so far been able to obtain quantitative
results for realistic systems such as Helium 4. 
One of the main difficulties is that new tools are needed to treat glassiness for quantum interacting 
particles systems in a consistent way (the majority of previous works focused on spin systems and
coarse-grained field theories).
Recent mean field studies based on Quantum Mode Coupling Theory (QMCT)~\cite{QMCT}
and on lattice models~\cite{FSZ11} gave new insights, and
a coherent qualitative picture of the glass transition in quantum
hard spheres has now emerged~\cite{FSZ11,Za11}.
The result is somehow surprising.
Basic intuition suggests that increasing the
strength of quantum fluctuations should
enable atoms to better explore the phase
space through tunneling, and thereby
inhibit the onset of dynamical arrest that is
necessary for glass formation. In contrast,
QMCT predicts~\cite{QMCT} that increasing
quantum fluctuations, by
augmenting the thermal wavelength of the
particles (\eg by reducing their mass), actually
favors the glass phase. Indeed, one finds that 
the glass phase emerges at lower density. This result has also been obtained in the exact solution
of mean field quantum lattice glass models~\cite{FSZ11}, and is corroborated
by path integral numerical
simulations~\cite{QMCT}.
The 
analysis of a toy model provided the following intuitive explanation~\cite{FSZ10}: The classical phase space
of a dense particle system is split into many different basins;
in the liquid phase the system visits a lot of basins that are
not very efficiently packed and therefore have a smaller free volume
(the volume accessible to each particle to vibrate around its equilibrium
position).
In the glass phase, the system instead visits a few rare basins
that are very efficiently packed and therefore have a larger free volume.
Adding quantum fluctuations, particles can lower their kinetic energy by delocalizing,
therefore the kinetic energy gain is proportional to free volume: hence, glassy states
have lower kinetic energy and are favored by quantum fluctuations.
The net result is that the glass transition line is re-entrant,
moving at lower density on increasing quantum fluctuations~\cite{QMCT}. 
For strongly quantum systems, however, the transition changes nature and becomes 
a first order phase transition between a superfluid phase and the glass~\cite{FSZ11}.
This first order transition is accompanied by phase coexistence, which leads to an 
heterogeneous ``superglass'', where regions of low-density
superfluid liquid coexist with regions of a high density glass. 
See~\cite{FSZ11,Za11} for a more detailed discussion.

The results mentioned above have been obtained on simplified models, therefore they 
only provide qualitative guideline for the physical behavior of Helium 4.
Quantitative analytic calculations of the phase boundaries for systems such as Helium 4 are still missing.
Unfortunately, Mode-Coupling Theory 
is of limited use since it is known to give poor estimates of the transition point and, at least 
at the present stage, neglects exchange effects~\cite{QMCT}.
The aim of this paper is to adapt the replica theory of glasses~\cite{Mo95,MP96,MP99,MP09,PZ10},
which has been shown to give good quantitative estimates of the glass transition point for classical liquids,
to the quantum case; and then apply it to Helium 4.

The paper is organized as follows. First, in section~\ref{sec:semi},
we use the standard prescription of replica theory to derive an expression of the
glass free energy in the semiclassical region. We show that this is enough
to obtain a quantitative estimate of the re-entrance effect at high density. 
In section~\ref{sec:phasediagram} we speculate on the possible phase diagram
of Helium 4 in the high density amorphous metastable region.
Next, in section~\ref{sec:discussion}, we discuss the assumptions we made 
and identify the main problems of the theory: we show that there
are some important conceptual difficulties that prevent us from extending the calculation
to the strongly quantum regime and from including exchange. While we believe that these difficulties can be
in principle overcome, we leave this task for future work.

\section{The semiclassical replica method}
\label{sec:semi}

We start our discussion by a straighforward application of the classical prescription of the
replica method to the quantum case. We consider a $d$-dimensional 
system of quantum Bosonic particles, of mass $M$,
interacting via a potential $v(r)$. As usual $h$ is the Planck's constant and $\hbar = h/2\pi$.
We set the Boltzmann constant $k_B=1$.
If $x_i$ is the position of particle $i$, with $i = 1,\cdots, N$,
the Hamiltonian is
\beq\label{Hsingle}
H = -\frac{\hbar^2}{2M} \sum_{i} \nabla_{x_{i}}^2 
+ \sum_{i<j} v(x_{i}-x_{j}) \ ,
\eeq
where $\nabla_{x_{i}}$ is the gradient with respect to $x_i$.

In the following we will make an extensive use of the  replica theory of classical glasses.
We refer to the original works~\cite{Mo95,MP99} and the review \cite{MP09} for details. 
The prescription of the theory
is to consider a system made of $m$ copies of the original system, subjected to an inter-replica
attractive coupling that we choose to be harmonic~\cite{MP99}. 
We denote by $x_{ai}$, $a = 1, \cdots, m$, the copies of particle $i$.
The Hamiltonian of such replicated
system is
\beq\label{Hcoupled}
H = -\frac{\hbar^2}{2M} \sum_{ai} \nabla_{x_{ai}}^2 
+ \sum_{a,i<j} v(x_{ai}-x_{aj}) + \frac{\ee}{2m}\sum_{i,a<b} (x_{ai}-x_{bi})^2 \ ,
\eeq
where, compared to~\cite{MP99}, $\ee$ has been rescaled by $m$ for later convenience.
Since we want to investigate the semiclassical limit, we neglect exchange for the moment,
and consider the particles as distinguishable.

\subsection{Small cage expansion}

For each molecule $i$ (we now omit the index $i$)
we consider the column vector $x$ such that $x^T = (x_1,\cdots, x_m)$.
We perform the change of variables $y_a = v_a^T x$ with
$v_1^T = \frac1{\sqrt{m}} (1,\cdots,1)$,
$v_a^T v_1 = 0$ and $v_a^T v_b = \d_{ab}$. The Jacobian
of the transformation is 1 because the vectors $v_a$ are orthonormal.
Thus the kinetic term is invariant under the transformation,
$ \sum_{a} \nabla_{x_{a}}^2 =  \sum_{a} 
\nabla_{y_{a}}^2$.
Defining the matrix of all unit entries, $1_{ab} = 1$, 
denoting by $\d$ the identity matrix,
and using the completeness relation $\d = \sum_a v_a v^T_a$, we can write
$\sum_{a<b} (x_a-x_b)^2 = 
x^T (m \d - 1) x = \sum_{ab} x^T v_a v^T_a (m \d -1) v_b v^T_b x =
\sum_{ab} y_a v^T_a (m \d -1) v_b y_b \ $.
It is easy to show that $(m \d -1) v_1 = 0$ and that 
$(m \d -1) v_a = m v_a$ for $a>1$. Then,
$\sum_{a<b} (x_a-x_b)^2 = \sum_{a=2}^{m} m y_a^2 \ $.
Thus the Hamiltonian in the $y$ variables reads
\beq
H = -\frac{\hbar^2}{2M} \sum_{ai} \nabla_{y_{ai}}^2 
+ \sum_{a,i<j} v(x_{ai}-x_{aj}) + \frac\ee2 \sum_{i,a>1} y_{ai}^2 \ ,
\eeq
where for each molecule $i$
\beq\label{xy2}
x_a = \sum_b (v_b)_a y_b = \frac{y_1}{\sqrt{m}} + \sum_{b>1}  (v_b)_a y_b 
= X + \sum_{b>1}  (v_b)_a y_b \ ,
\eeq
with $X = y_1/\sqrt{m} = (\sum_a x_a)/m$ the center of mass of the molecule.

Now, following \cite{MP99}, we perform a large $\ee$ expansion.
For large $\ee$ we assume that the $y_a$, $a>1$ are small due to the harmonic term,
and we expand the
potential $v(x_{ai}-x_{aj})$. 
For a given pair of indices $i, j$ (that we now omit),
we define $\D x_a = x_{ai}-x_{aj}$, and it follows from Eq.~(\ref{xy2}) that
$\D x_a = \D X + \sum_{b>1}  (v_b)_a \D y_b$.
We denote by $\mu, \nu = 1, \cdots, d$ the
spatial indices, and $\partial_\m v = dv/d x^\mu$ 
the derivative with respect to coordinate $x^\mu$.
Then we can easily expand the potential as follows:
\beq\begin{split}
\sum_a v(\D x_a) &  = m v(\D X) + \sum_a \partial_\m v (\D X) \sum_{b>1} (v_b)_a \D y^\m_b
\\ &+\frac12 \sum_a \partial_{\m\n} v(\D X) 
\sum_{b>1} (v_b)_a \D y^\m_b\sum_{c>1} (v_c)_a \D y^\n_c \ .
\end{split}\eeq
Recalling that for $b>1$, $\sum_a (v_b)_a = \sqrt{m} v_1^T v_b = 0$ and that
$\sum_a (v_b)_a(v_c)_a = v_b^T v_c = \d_{bc}$, we get
$
\sum_a v(\D x_a)= m v(\D X) + \frac12  \partial_{\m\n} v(\D X) \sum_{b>1} \D y^\m_b \D y^\n_b \ .
$
Collecting the terms together and changing variables to $X_i = y_{1i}/\sqrt{m}$,
we finally get
\beq\label{HXy}
\begin{split}
H & = \underbrace{ -\frac{\hbar^2}{2mM} \sum_{i} \nabla_{X_{i}}^2  + \sum_{i<j} m v(X_i-X_j)}_{H_0(X)}
+ \sum_{i,a>1,\mu} \underbrace{\left[ - \frac{\hbar^2}{2 M} \nabla_{y^\m_{ai}}^2
+ \frac\ee2 ( y^\mu_{ai})^2 \right]}_{H_{\rm harm}(y^\mu_{ai})} \\
& + \underbrace{ \frac12\sum_{i<j} \partial_{\m\n} v(X_i-X_j) \sum_{a>1} (y_i-y_j)^\m_a (y_i-y_j)^\n_a }_{V_{\rm int}(X,y)} 
\ .
\end{split}\eeq
At this order, the Hamiltonian is the sum of a term $H_0(X)$ that describes the system
of the centers of mass, with a renormalized mass $mM$ and renormalized interaction $m v(r)$,
plus a sum of independent terms $H_{\rm harm}(y)$, 
that describe harmonic oscillators of spring constant $\ee$. 
The centers of mass are
coupled to vibrations by the term $V_{\rm int}$; since the vibrations are small for large $\ee$, 
this coupling is small and we can treat it as a perturbation.

\subsection{Computation of the partition function}

We now want to compute the partition function of the system at temperature $T$,
using the Hamiltonian in (\ref{HXy}). Again, we stress that in the following we consider that
the particles are distinguishable. Otherwise, even in the absence of the coupling $V_{\rm int}$,
one should consider the $y$ variables as ``internal'' degrees of freedom (like spins) 
of the centers of mass, and one should therefore impose the permutation symmetry on the global
wavefunction, that would correlate the $X$ and $y$ parts of the wavefunctions.
Instead, for distinguishable particles, there is no special symmetry requirement and 
we consider that in absence of $V_{\rm int}$ the space of wavefunctions is a product
of independent wavefunctions for the $X$ and $y$ parts.
Under these assumptions, we get, developing at first order in $V_{\rm int}$:
\beq\nonumber
Z_m(\ee) = \Tr \, e^{-\b H} = 
\Tr_X \, e^{-\b H_0(X)} 
\left(\Tr_y \, e^{-\b H_{\rm harm}(y)} \right)^{N d (m-1)} 
 \big[ 1 - \b \la V_{\rm int} \ra + O(V_{\rm int}^2) \big] \ ,
\eeq
where $\la V_{\rm int} \ra$ is the quantum and thermal average of the perturbation over the $X$ and $y$.
The partition function and free energy of each independent oscillator is
\beq\begin{split}\label{Fosc}
& Z_{\rm harm}(\ee) = \Tr_y e^{-\b H_{\rm harm}(y)} = \frac{e^{\b\hbar \om/2}}{e^{\b\hbar\om}-1} \ , \\
& F_{\rm harm}(\ee) = -T \log  Z_{\rm harm}(\ee) =
-T \log \frac{e^{\b\hbar \om/2}}{e^{\b\hbar\om}-1}
= U_{\rm harm}(\ee) - T S_{\rm harm}(\ee)
 \ ,
\end{split}\eeq
where the frequency $\om = \sqrt{\ee/M}$; there are $d (m-1)$ independent 
oscillators per molecule.
It will also be useful to define the mean square displacement of the oscillator:
\beq\begin{split}
& \la y^2 \ra_{\ee} = 2 \frac{dF_{\rm harm}(\ee)}{d\ee} =  
\frac{\hbar\om}{\ee}  \left[ \frac12 + \frac1{e^{\b \hbar \om}-1} \right]  = 
\frac{\hbar\om}{\ee}\bar n(\om) \ , \\
\end{split}\eeq
with $\bar n(\om) =  1/2 + 1/(e^{\b \hbar \om}-1) \ $.

The average of the interaction term over $y$ is
\beq\begin{split}
\la V_{\rm int} \ra_y &= 
\frac12  \sum_{i<j}  \partial_{\m\n} v(X_i-X_j)
\sum_{a>1} \la (y_i-y_j)_a^\m (y_i-y_j)_a^\n \ra_y \\
&= (m-1) \la y^2 \ra_{\ee} \sum_{i<j} \Delta v(X_i-X_j) \ ,
\end{split}
\eeq
using that different spatial components and sites $i$ and $j$ are uncorrelated,
$\la (y_i-y_j)_a^\m (y_i-y_j)_a^\n \ra_y =
\d_{\m\n} \la ( y_{ia}^\m)^2 + ( y_{ja}^\m)^2  \ra_y 
= 2 \d_{\m\n} \la y^2 \ra_{\ee}$.
Taking the average of $V_{\rm int}$ also on the centers of mass, we 
get the first order correction to the free energy:
\beq\label{Fmeps}
\begin{split}
\Phi(m,\ee) & = -\frac{T}{N} \log Z_m(\ee) =  
 - T d (m-1) \log \left[ \frac{e^{\b\hbar \om/2}}{e^{\b\hbar\om}-1}\right] \\
& -\frac{T}N \log \Tr_X e^{-\b H_0(X)} 
+ (m-1) \la y^2 \ra_{\ee} \left\la \frac1N \sum_{i<j} \Delta v(X_i-X_j) \right\ra_0 \\
 = & d (m-1)  F_{\rm harm}(\ee)
+ F_0(m) + 
(m-1) \la y^2 \ra_{\ee} \frac{\r}2 \int dr \Delta v(r) g_0(r)
\ ,
\end{split}\eeq
where $F_0(m) = -T/N \log \Tr_X e^{-\b H_0(X)}$ is the free energy of the effective liquid
of the centers of mass,
and
$\r^2 g_0(X-Y) = \la \sum_{i\neq j} \d(X-X_i) \d(Y-Y_j) \ra_0$
is its pair correlation function.
In the following we will use the notation $\la \Delta v(r) \ra=\frac\r{2} \int dr \Delta v(r) g_0(r)$.

Note that the Hamiltonian $H_0(X)$ can be rewritten as
$H_0(X)=m \big[ -\sum_i \frac{\hbar^2}{2 m^2 M} \nabla_{X_i}^2 + \sum_{i<j} v(X_i-X_j) \big]$,
then the fluid of centers of mass can be viewed as the original fluid at inverse temperature
$\wt\b = m \b$ and with a mass $\wt M = m^2 M$.
Therefore:
\beq
\frac{1}{m} F_0(m) =  -\frac{T}{mN} \log \Tr_X e^{-\b H_0(X)} = F_{\rm liq}(m\b,m^2 M) \ ,
\eeq
where $F_{\rm liq}(\b,M)$ is the free energy associated to the original Hamiltonian (\ref{Hsingle}).

\subsection{The replicated free energy}

Recalling that $\frac{d F_{\rm harm}(\ee)}{d\ee} = \frac12 \la y^2 \ra_{\ee}$, we rewrite 
(\ref{Fmeps}) as
\beq\label{Freseps}
\begin{split}
\Phi(m,\ee) & =  m F_{\rm liq}(m\b,m^2 M) + d (m-1) \left[ F_{\rm harm}(\ee) + \frac{d F_{\rm harm}(\ee)}{d\ee} \frac{2\la \D v(r) \ra}d
\right] \\ &\sim
 m F_{\rm liq}(m\b,m^2 M) + d (m-1) F_{\rm harm}(\ee + \k ) 
\ ,
\end{split}\eeq
having defined the effective spring constant $\k =2 \la \D v(r) \ra / d$.
This is correct as long as $\k \ll \ee$, which is indeed the case since we are working in
the limit of large $\ee$. 
Now, following \cite{MP99}, we have to extrapolate the free energy obtained in the large $\ee$ expansion 
to the limit of zero coupling $\ee$. This can be done via a Legendre transformation with respect to $\ee$,
following closely the classical derivation of~\cite{MP99}. This procedure provides a formally correct treatment that shows
that the large $\ee$ expansion is equivalent to a small cage expansion~\cite{MP99}.
However, we see that in Eq.~(\ref{Freseps}) can set directly $\ee=0$ and we obtain a
meaningful result for the replicated free energy, which indeed coincides with the one obtained via the Legendre
transform procedure. For the sake of coinciseness, we adopt this procedure here: 
see~\cite{MP99} for further details on the small cage expansion procedure.
Finally, we get for the repicated free energy:
\beq\label{Phim}
\Phi(m) =  m F_{\rm liq}(m\b,m^2 M) + d (m-1) F_{\rm harm}(\k) \ ,
\eeq
where the function $F_{\rm harm}(\ee)$ 
is defined in (\ref{Fosc}) and $\k =2 \la \D v(r) \ra_{\rm liq}/d$, 
where the average is over the liquid
with temperature $T/m$ and mass $m^2 M$.

Given $\Phi(m)$, the thermodynamic properties of the glass,
\ie the internal energy $f(\b,m)$ and the complexity $\Si(\b,m)$,
are obtained following~\cite{Mo95,MP99}:
\beq\begin{split}
&f(\b,m) = \frac{\partial \Phi(m)}{\partial m} \ , 
\hskip2cm
\Si(\b,m) = \b m^2 \frac{\partial \Phi(m)/m}{\partial m} \ .
\end{split}\eeq
Using the relations
$S_{\rm liq}(T)=  -\partial_T F_{\rm liq}(T,M)$ (entropy), 
$K_{\rm liq}(T) = -M \partial_M F_{\rm liq}(T,M)$ (kinetic energy), we get
for the equilibrium complexity (\ie the complexity at $m=1$):
\beq\label{SiT}
\begin{split}
\Si_{\rm eq}(T) & = \Si(\b,1) = S_{\rm liq}(T) - 2\b K_{\rm liq}(T) + \b d \, F_{\rm harm}(\k) 
\\ &= 
S_{\rm liq}(T) - d \, S_{\rm harm}(\k) - \b [ 2 K_{\rm liq}(T) -d \, U_{\rm harm}(\k) ] \ .
\end{split}\eeq
In the classical case the last term vanishes due to equipartition, since
$U_{\rm harm} = T$ and $K_{\rm liq} = d T/2$.
This is not {\it a priori} true in the quantum regime, but it should be
true if we assume that the liquid is a superposition of ``glassy''
metastable states corresponding to harmonic vibrations around 
amorphous positions.  We will comment later on this crucial issue.

\subsection{The classical limit}

Let us first consider the classical limit $\hbar \to 0$. 
We have $\bar n(\om) \to (\b\hbar\om)^{-1}$,
and 
\beq
F_{\rm harm}(\ee) \to T \log \b \hbar \om = T \log \L +\frac{T}2 \log \frac{\ee}{2\pi T} \ ,
\eeq
where $\L = h/\sqrt{2\pi M T}$ is the thermal wavelength.
Moreover the liquid of the centers of mass is
classical and has a partition function
\beq
Z_{\rm liq} = \int \frac{d^N P d^N X}{h^N N!} e^{-\frac{\b}{m} \sum_i \frac{P_i^2}{2M} - \b m \sum_{i<j} v(X_i-X_j)}
= \L^{-d N} m^{d N/2} Z_c(\b m) \ ,
\eeq
where $Z_c(\b) = (N!)^{-1}\int d^N X e^{-\b \sum_{i<j} v(X_i-X_j)}$ is the configurational partition function. 
Therefore,
$F_0(m) = m F_{\rm liq}(m\b,m^2 M) = d \, T \log \L - \frac{d \, T}2 \log m + m F_c(\b m)$, 
where $F_c(\b) = -(T/N) \log Z_c(\b)$.
Using these results, Eqs.~(\ref{Phim}) and (\ref{SiT}) become
\beq\label{SiEqClassical}
\begin{split}
\Phi(m)& = d \, m \, T \log\L -\frac{d \, T}{2}\log m + m F_c(\b m) + \frac{d(m-1)T}{2} 
\log \frac{\la \Delta v(r) \ra^*}{\pi T d} \ , \\
\Si_{\rm eq}(T) &= S_{\rm liq}(T) + d \log \L + \frac{d}2 \log  \frac{\la \Delta v(r) \ra^*}{\pi T d}
- d \\ &= 
S_c(T) + \frac{d}2 \log  \frac{\la \Delta v(r) \ra^*}{\pi T d}
- \frac{d}2 \ ,
\end{split}
\eeq
where $S_c(T) = S_{\rm liq}(T) + d \log\L - d/2$ and $\la \bullet \ra^*$ denotes an average 
over the classical liquid at temperature $T^* = T/m$ (with $m=1$ in the second line).
These are exactly the results obtained by M\'ezard and Parisi in the classical case~\cite{MP99}.

\subsection{First order quantum correction}

We now compute the first order quantum correction to the complexity 
in the semiclassical expansion.
This can be done using a formal expansion in powers of $\hbar$~\cite{hbarexp}.
For simplicity we will perform the computation directly for $m=1$,
starting from Eq.~(\ref{SiT}).
The correction to the free energy of the liquid is
\beq\label{dFq}
\d F_{\rm liq} = \frac{\L^2}{24 \pi} \la \Delta v(r) \ra^* \ ,
\eeq
then it is easy to see that
\beq
\d [  S_{\rm liq}(T) - 2\b K_{\rm liq}(T) ] = -\frac{\L^2 }{24 \pi} \left[ \frac1T \la \Delta v(r) \ra^*
+ \partial_T \la \Delta v(r) \ra^* \right] \ .
\eeq
To compute the variation of the free energy of the oscillator, we must take into
account that the spring constant $\k$ also has quantum corrections. From (\ref{dFq}) we get
$\d g_0(r) = \frac{\L^2}{24 \pi} \Delta g_0(r)$,
and using this we get
\beq
\frac{d}2 \, \d \k = \frac\r{2} \int dr \, \d g_0(r) \, \D v(r) =
\frac{\r}2 \int dr \frac{\L^2}{24\pi} \D g_0(r) \, \D v(r) = \frac{\L^2}{24 \pi} \la \D^2 v(r) \ra^* \ .
\eeq
Denoting $\k_0 = 2\la \Delta v(r) \ra^*/d$
and $\om_0 = \sqrt{\k_0/M}$ the classical spring constant and frequency,
the first correction to the free energy of the oscillator is:
\beq
\d [\b d F_{\rm harm}(\k)] = d \frac{(\b\hbar \om_0)^2}{24} + \frac{d}2 \frac{\d \k}{\k_0}
= \frac{\L^2}{24 \pi T} \la \Delta v(r) \ra^* + 
\frac{d \, \L^2}{48 \pi}\frac{ \la \D^2 v(r) \ra^*}{\la \Delta v(r) \ra^*} \ .
\eeq
Collecting these results we obtain
\beq\label{SiEqQuantum}
\d \Si(T) = \frac{ \L^2}{48 \pi}\left[ d\frac{ \la \D^2 v(r) \ra^*}{\la \Delta v(r) \ra^*}
- 2 \, \partial_T \la \Delta v(r) \ra^* \right] \ .
\eeq
Note that using the results above one can check that 
at this order in $\hbar$ we still have $2 K_{\rm liq} = d \, U_{\rm harm}(\k)$ so the
last term in (\ref{SiT}) is still zero.

\subsection{Results}

\begin{figure}
\begin{center}
\includegraphics[width=.48\textwidth]{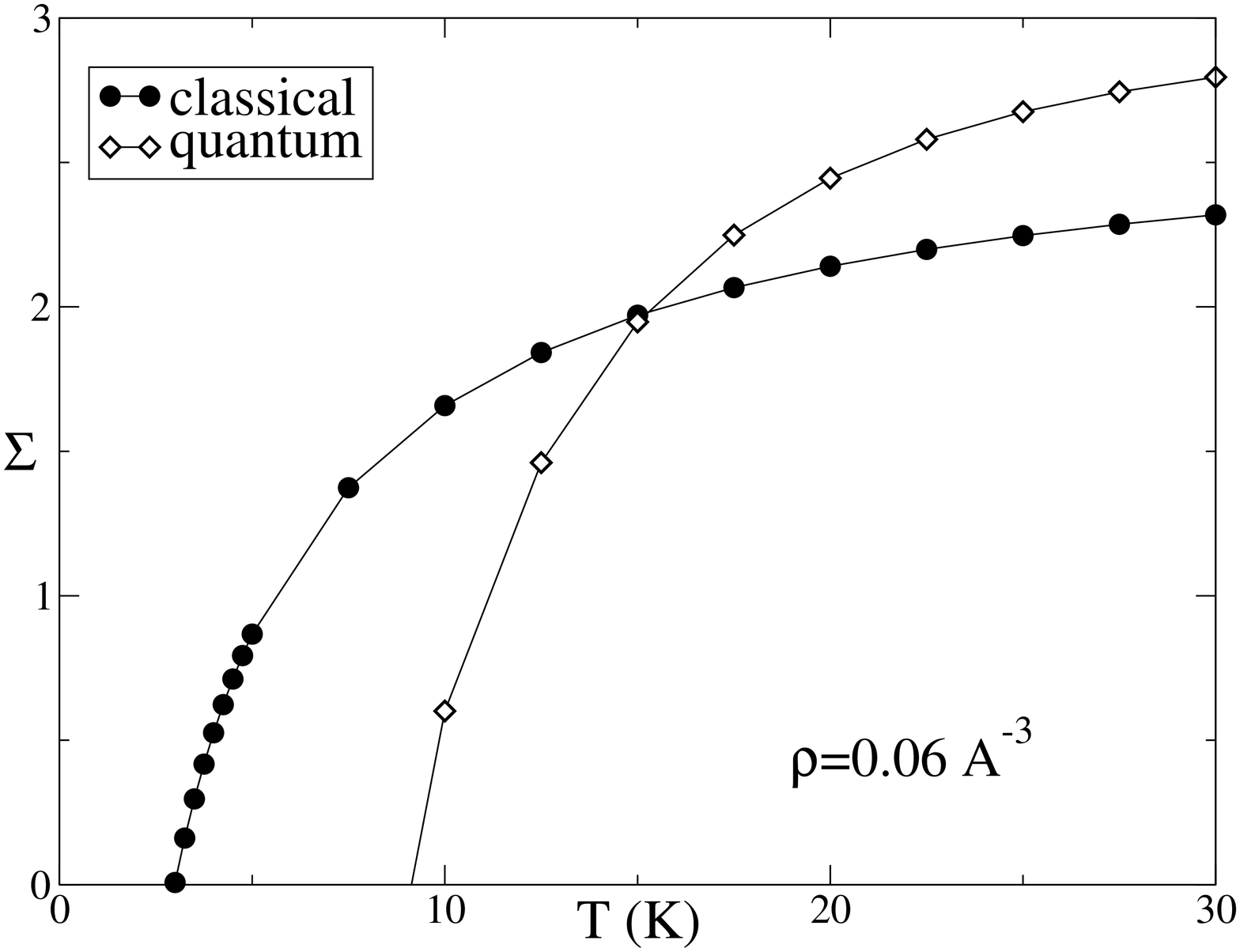}
\includegraphics[width=.48\textwidth]{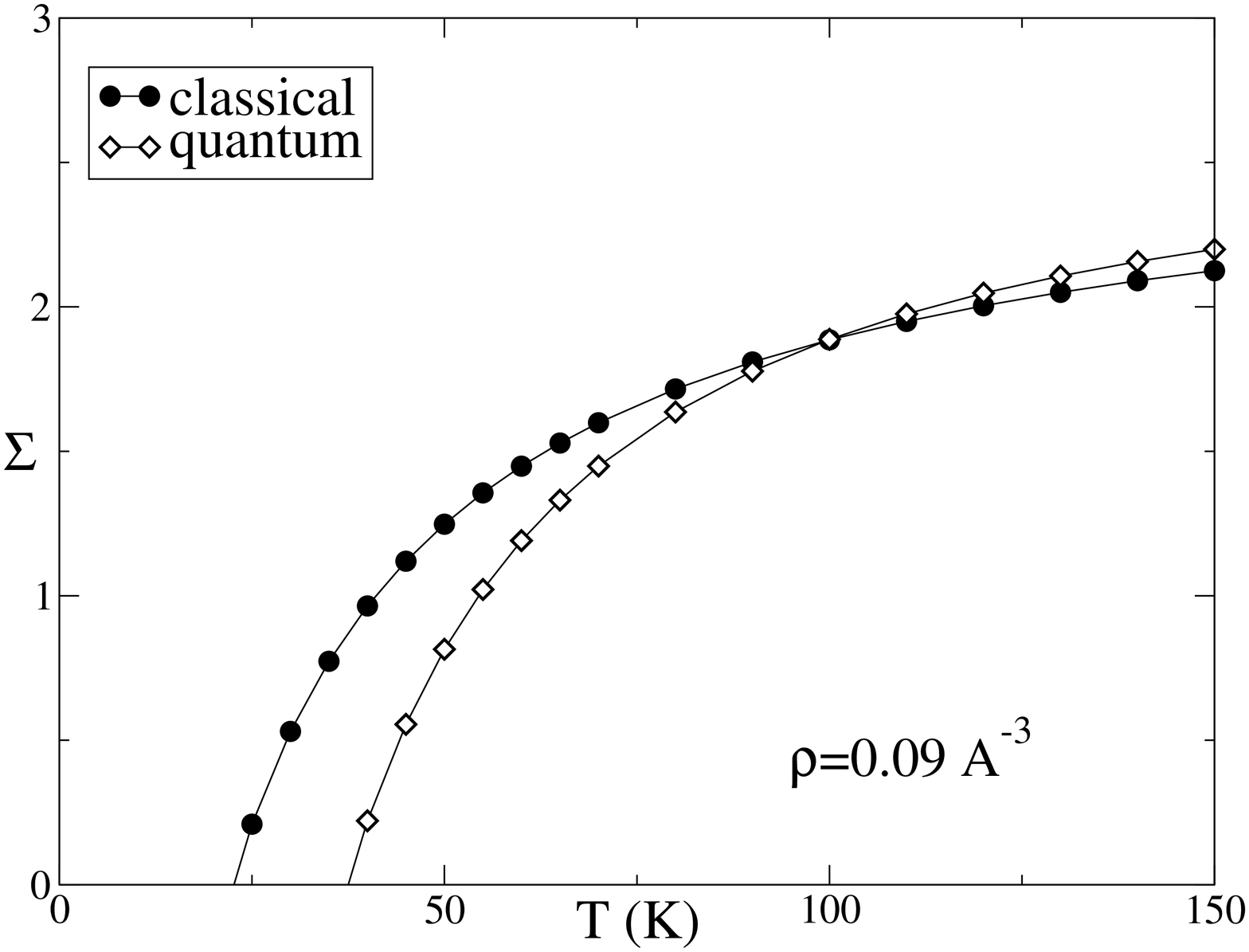}
\end{center}
\caption{
The complexity $\Si_{\rm eq}(T)$, both in the classical
limit, Eq.~(\ref{SiEqClassical}), and with the quantum 
correction, Eq.~(\ref{SiEqQuantum}), for two representative densities
as a function of temperature.
}
\label{fig:sigma}
\end{figure}

We now present and discuss the results of the semiclassical expansion presented
above. Note that altough Eq.~(\ref{SiEqQuantum}) has been derived in a formal expansion
in powers of $\hbar$, we will in the following apply it using
the real values of all physical constants.
We focus on Helium 4, which we describe by means of a Lennard-Jones (LJ) potential
$v(r) = 4 \ee \big[ \big( \frac{\s}r \big)^{12} - \big( \frac{\s}r \big)^6 \big]$
with $\ee = 10.22$ K and $\s = 2.556 \, \AA$~\cite{Ce95}. Given the Helium 4 mass,
we have $\hbar^2/(2 M) = 6.0596$ K $\AA^2$. The intensity of quantum fluctuations can be
conveniently quantified by the ratio of the thermal
wavelength and the inter-particle distance,
$\L^* = \L \r^{1/3} = \r^{1/3} \sqrt{4 \pi \hbar^2/( 2 M T )} = 
8.726 \, \AA \, \text{K}^{1/2} \times \r^{1/3} T^{-1/2}$ .
Although better potentials
have been constructed~\cite{Ce95}, we choose the Lennard-Jones because it is the
simplest and it gives reasonable quantitative results; in addition, the glass transition
of a classical monoatomic Lennard-Jones system has been studied in great detail both 
analytically~\cite{MP99} and numerically~\cite{DiLeonardo00}, so we will be able to
compare the results obtained in the classical case with the first order quantum correction.

To compute the classical equilibrium complexity and its quantum correction,
according to Eqs.~(\ref{SiEqClassical}) and (\ref{SiEqQuantum}), we need
the entropy of the classical liquid, as well as its $g(r)$, 
as a function of temperature. Note that since $m=1$, the averages $\la \bullet \ra^*$
reduce to standard thermal averages over the classical liquid, that can be computed
using $g(r)$. These quantities can be obtained using one of the integral equations of 
classical liquid theory~\cite{Hansen}. Following \cite{MP99},
we choose the HNC closure~\cite{Hansen} which has the advantage of providing a direct
route to compute the free energy, and gives thermodynamic quantities with reasonable
accuracy (of the order of $10\%$). The HNC equations for the LJ potential have been
solved numerically using standard iteration schemes.

In Fig.~\ref{fig:sigma} we report the complexity $\Si_{\rm eq}(T)$ (also called configurational
entropy in the context of structural glasses).
The point where it vanishes marks the Kauzmann transition
$T_{\rm K}$ to the ideal glass.
We see that although the quantum correction to $\Si_{\rm eq}(T)$ can be positive or negative,
close to $T_{\rm K}$ it is always negative, therefore
increasing the value of $T_{\rm K}$ with respect to the classical limit.
At the lowest densities, $T_{\rm K}$ is found to increase by a factor of two with
respect to the classical value. It is worth to note, however, that the semiclassical
expansion is particularly bad for $\rho \leq 0.055 \, \AA^{-3}$: in this region,
the classical Lennard-Jones system enters the gas-liquid
phase separation region~\cite{Smit}, and the HNC equations cease to converge.
Instead, for quantum Helium 4 the phase coexistence region
is known to be much smaller, ending at $\rho \sim 0.022 \, \AA^{-3}$~\cite{CSB96}. This
is due to the large zero-point kinetic energy, and it signals that a semiclassical expansion
is definitely not reasonable at low density, as expected.

In Fig.~\ref{fig:PD_He} (left panel) we report the glass transition temperature $T_{\rm K}(\rho)$, obtained
from the classical and the quantum computation. We see that at all densities, the quantum value
is larger than the classical one (with the relative difference decreasing on increasing $\rho$).
The classical value compares well with the value obtained from a numerical simulation of the
classical LJ fluid~\cite{DiLeonardo00}. The quantum value is below the melting temperature
of Helium 4, as it should. Unfortunately, as already discussed, we cannot obtain results for the
really interesting region of low density and temperature, where quantum effects are most important.
For the classical LJ 
potential, $T_{\rm K}$ increases with density, therefore the transition line
in the $(\r , \L)$ plane is re-entrant. This is a trivial classical effect and has nothing to
do with the re-entrance found in \cite{QMCT,FSZ10,FSZ11}. The important point, for the LJ case,
is that quantum corrections increase the value of $T_{\rm K}$,
thereby promoting glass formation as stated in \cite{QMCT,FSZ10,FSZ11}.
The region $\L^* \sim 1$, where the semiclassical computation breaks down,
is where we expect that the system should undergo a phase transition towards a 
superfluid state (the line $\L^* = 1$ is reported in Fig.~\ref{fig:PD_He} for comparison).
Since exchange was neglected from the very beginning, we cannot obviously analyze this transition.

\section{Speculations on the phase diagram of high density metastable Helium 4}
\label{sec:phasediagram}

\begin{figure}
\begin{center}
\includegraphics[width=.48\textwidth]{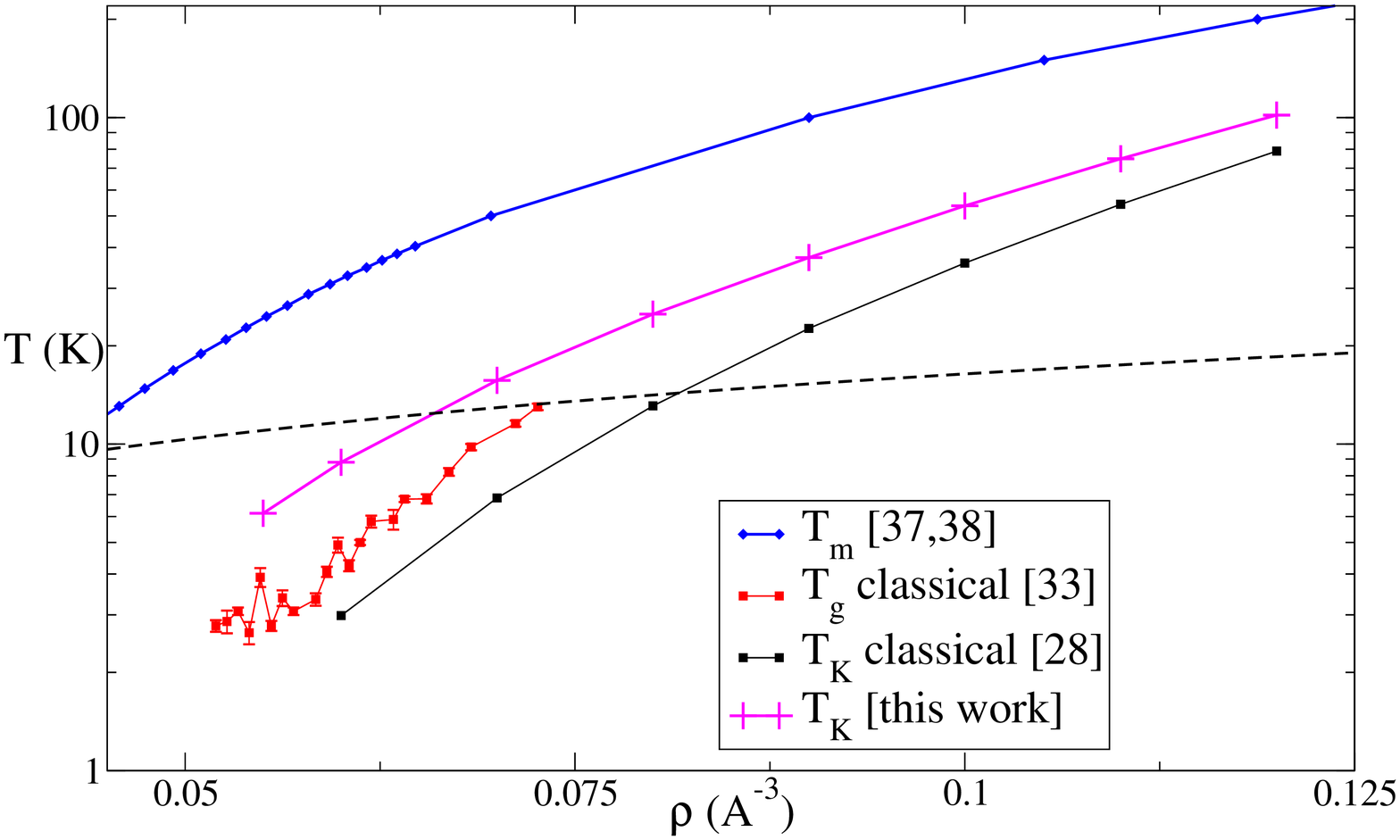}
\includegraphics[width=.48\textwidth]{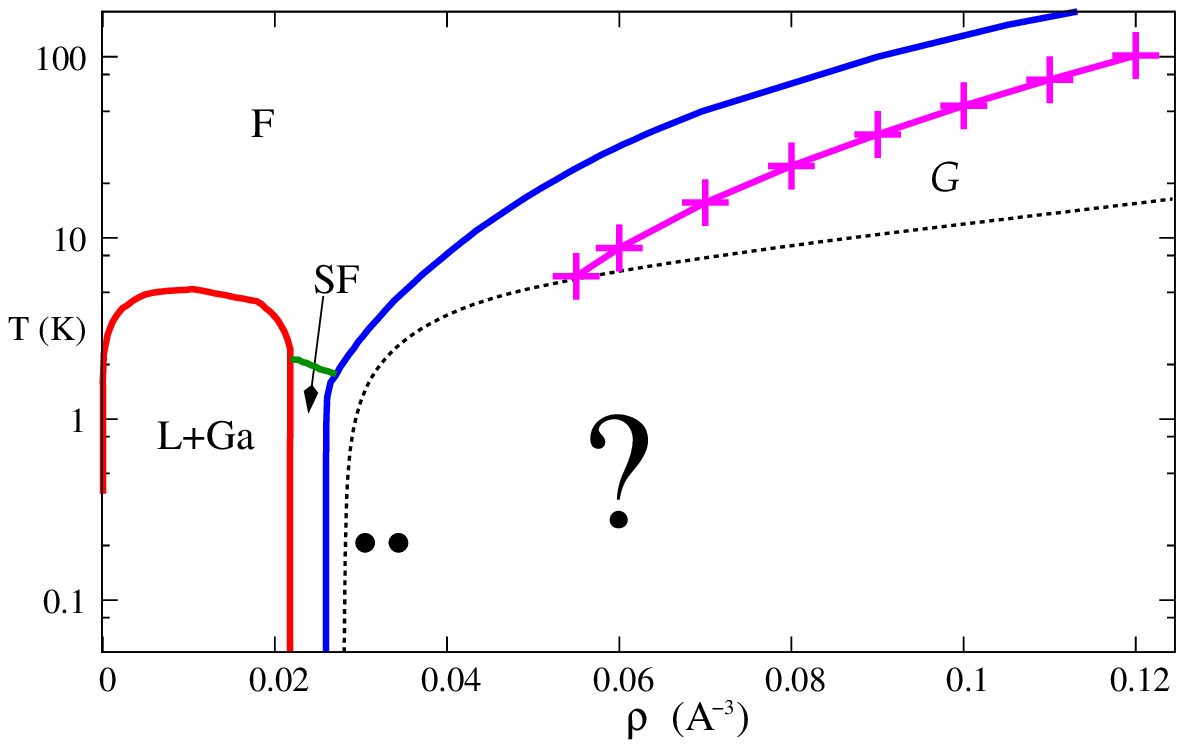}
\end{center}
\caption{(Color online)
({\it Left panel}) Glass transition lines $T_{\rm K}(\r)$, defined by $\Si_{\rm eq}(T)=0$, as obtained
from the classical limit Eq.~(\ref{SiEqClassical})~\cite{MP99} and including the first semiclassical quantum 
correction Eq.~(\ref{SiEqQuantum}).
Also shown is a numerical determination of the 
classical glass transition $T_{\rm g}(\r)$ from Ref.~\cite{DiLeonardo00}, and the numerically and experimentally determined  
melting temperature $T_{\rm m}(\rho)$ of Helium 4, obtained from Refs.~\cite{CSB96,YMR81}.
The line $\L^* = \L \r^{1/3} = 1$ below which quantum effects are very strong is plotted as a dashed line.
({\it Right panel})
Complete temperature-density phase diagram of Helium 4.
{\it Equilibrium phases} (from Refs.~\cite{CSB96,YMR81}): 
The red line delimits the liquid (L)-gas (Ga) coexistence region. The green
line is the $\l$-line that separates the liquid from the superfluid (SF).
The blue line is the melting transition below which the crystal phase
is stable, and the disordered phase is metastable. 
{\it Glass phase:}
The purple line is the glass (G) transition line obtained in
the present work within the first order semiclassical expansion.
The behavior of the amorphous phase, say, within the dashed black line
is currently not well understood. The two black dots mark the state point
investigated numerically in Ref.~\cite{BPS1}.
}
\label{fig:PDboh}
\label{fig:PD_He}
\end{figure}

In Fig.~\ref{fig:PDboh} (right panel) 
we report in the density-temperature plane\footnote{The reason why we do not discuss the phase diagram in the more common pressure-temperature plane is that
the computation of the pressure from the semiclassical expansion would introduce additional uncertainty
on the position of the glass transition line. Moreover, both theoretical and numerical computations are
more naturally done at constant density.} all the information now
available for the phase diagram of Helium 4. We also draw the glass transition line that can be reached 
if crystallization can be avoided, see \eg \cite{supercooled} for an experimental investigation of supercooled 
liquid Helium. Ref.~\cite{BPS1} reports the existence of 
a metastable disordered superfluid state at $T=0.2$ K and densities $\r = 0.0292 \, \AA^{-3}$ and
$\r = 0.0359 \, \AA^{-3}$.  These state points are marked by a black dot in Fig.~\ref{fig:PDboh}.
However, it is not clear to us whether this state is a liquid or a glass\footnote{
In Ref.~\cite{BPS1} the system was named a glass because it was observed that a non-homogeneous density profile 
lasted for a certain time during a Path Integral Monte Carlo (PIMC) simulation. However, although certainly very stimulating, 
we find that these results are not conclusive
because of the following reasons:
\begin{itemize}
\item PIMC simulations only give access to a fictitious classical dynamics of a system of polymers that represent the
imaginary time evolution of the quantum particles. Despite some claims that this dynamics might be close to the true Schr\"odinger
dynamics of quantum particles (see \eg~\cite{QMCT}), the connection between the two has never been convincingly shown.
\item We don't find the evidence reported in Ref.~\cite{BPS1} strong enough to support the existence of a glass phase, even for the PIMC polymers.
Our point is that, by analogy with simpler classical systems, in order to establish the presence of a glass phase one has to perform a careful systematic
study of time-dependent density correlations, and show that such correlations develop a plateau at intermediate time scales, that there is a clear separation
of time scales between a short time (so-called $\b$) and long time (so-called $\a$) relaxation, and that there is aging in the glass phase. In absence of these
evidences, avoided or interrupted nucleation of a polycrystalline phase can easily be confused with the formation of a true glass, as it has been shown in a number
of studies on classical systems. An example of this
in the context of distinguishable Helium 4 particles was given in Ref.~\cite{BCFZ11}.
\end{itemize}
For this reason, the system of Ref.~\cite{BPS1} might be in reality a moderately viscous supercooled liquid, and not a really arrested glass.
We hope that further numerical work will clarify this issue.
}.

As already stressed, our results for the glass (purple) transition line
are reliable only in the high density and high temperature regime in the $(\r,T)$ phase diagram, 
where a semiclassical
expansion makes sense. Since our theory is unable to include exchange correctly,
we cannot describe the interplay between superfluidity and glassiness taking place
at low density and low temperature, say, within the dashed
black line in the phase diagram of Fig.~\ref{fig:PDboh}.
In the rest of this section we describe two possible scenarii for the transitions taking place in the region
corresponding to the question mark.

\subsection{A first order superfluid transition: the heterogeneous superglass phase}

\begin{figure}
\begin{center}
\includegraphics[width=.48\textwidth]{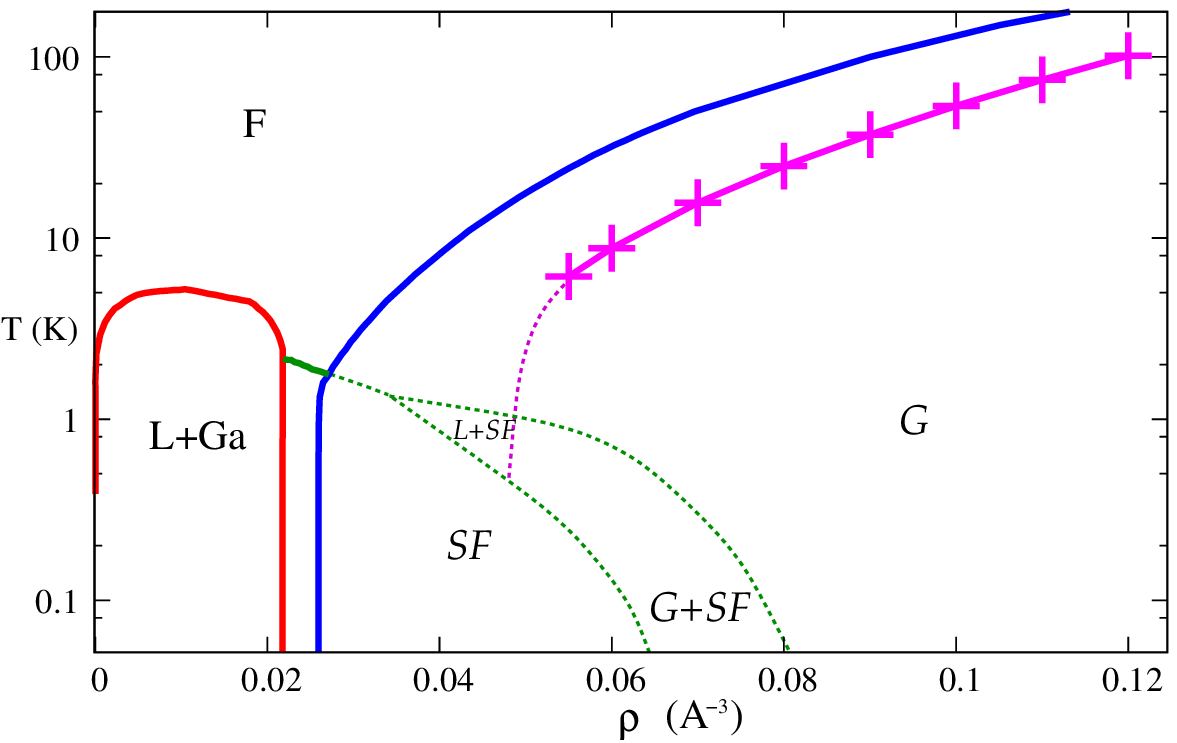}
\includegraphics[width=.48\textwidth]{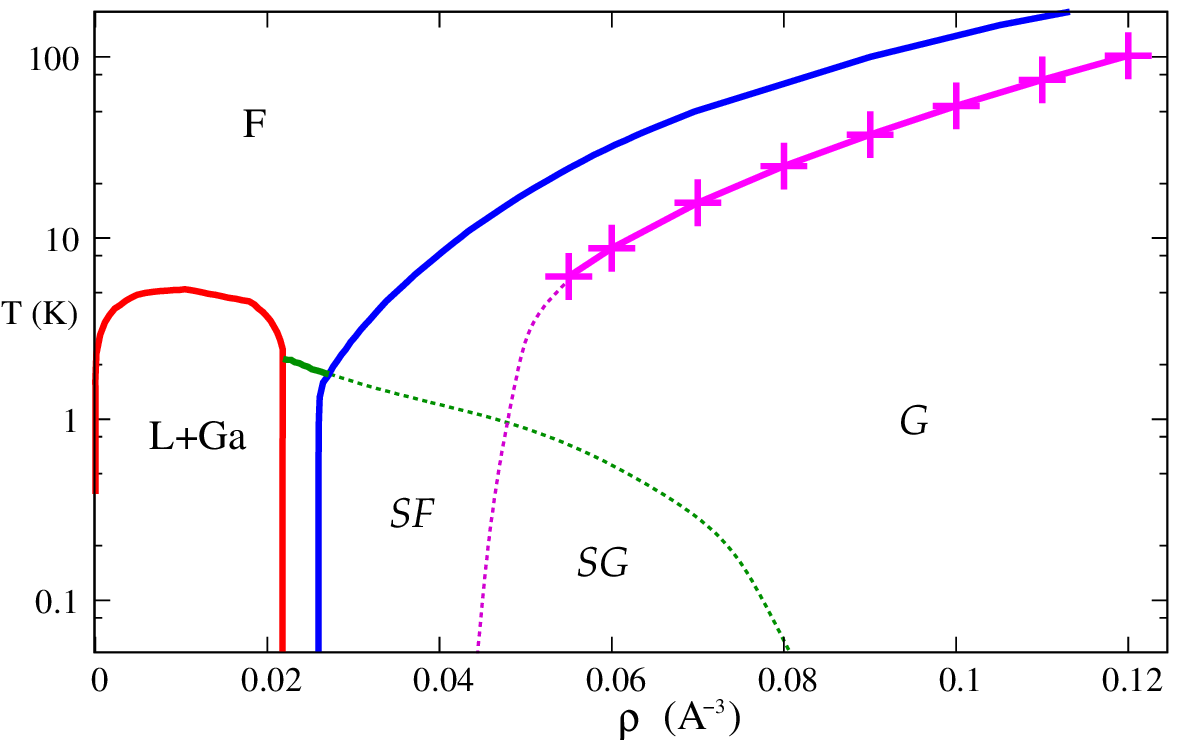}
\end{center}
\caption{(Color online)
({\it Left panel}) A conjectural phase diagram for metastable Helium 4 as it emerges from 
the combination of Refs.~\cite{QMCT,FSZ11,Za11} and the present study.
The stable part of the phase diagram is as in Fig.~\ref{fig:PDboh}.
The metastable phases are indicated in italic.
The purple line is the glass transition line obtained in
the present work within the first order semiclassical expansion.
The $\l$-line
becomes, at some point inside the supercooled liquid, a first order transition
line~\cite{FSZ11}. One should therefore observe a
superfluid (SF)-glass (G) coexistence. 
Note that we currently do not have any quantitative
estimate of the behavior of the $\l$-line inside the metastable phase, which is drawn
at arbitrary position in the figure.
({\it Right panel}) Another possible phase diagram for metastable Helium 4.
Same colors and remarks as in the caption of Fig.~\ref{fig:PDHEcon}.
The $\l$-line always stays a second order transition. When it crosses the
glass transition line, a superglass (SG) phase appears.
}
\label{fig:PDHEcon}
\label{fig:PDHEcon2}
\end{figure}

In Fig.~\ref{fig:PDHEcon} (left panel) we report a conjectural phase diagram for metastable Helium 4 in
the $(\r,T)$ plane, as it emerges by the combination of Refs.~\cite{QMCT,FSZ11,Za11} and the present study.
The purple line is the glass transition line obtained in
the present work within the first order semiclassical expansion. Below it,
the system is a glass.
In~\cite{FSZ11} it was predicted that the $\l$-line might become a first order transition
line at high density. Although this prediction has been made for a quite abstract 
lattice mean field model of dense liquids, 
it is interesting to discuss it since these models in several cases display a phenomenology
that is very similar to particle systems.
One should then observe a superfluid (SF)-glass (G) coexistence, leading to an heterogeneous
superglass where regions of low-density superfluid 
would coexist with regions of high-density glass.
The reader should keep in mind that 
the qualitative behavior (in particular the form) of the $\l$-line in the metastable phase is entirely conjectural.
Moreover we currently don't have any estimate of the exact location of this line; therefore,
its position in Fig.~\ref{fig:PDHEcon} is arbitrary.

\subsection{The homogeneous superglass phase}

A second scenario is that the $\l$-line always continues to be a second order transition in the metastable part
of the phase diagram. In this case, it will very probably cross the glass transition line at some point, leading
to a {\it superglass} phase (SG). This picture, which is displayed in Fig.~\ref{fig:PDHEcon2} (right panel),
is consistent with two quite different physical scenarios for the
glass transition line at low temperature:
\begin{itemize}
\item
In Refs.~\cite{CTZ09,TGIGM10}, a phase diagram like the one in Fig.~\ref{fig:PDHEcon2} (right panel) was found.
However, in the models investigated there, the glass  transition is continuous, like for spin glasses:  
the amplitude of frozen-in density fluctuations changes continuously at the transition.
This is very different from the transition of structural glasses, that we actually find in our semiclassical
analysis, and that corresponds to a discontinuous jump of the amplitude of the frozen-in density fluctuations, 
the transition remaining however of second order. 
To match these two findings one can conjecture that below some temperature the glass transition line
ceases to be discontinuous and becomes continuous instead. If this is the case, at low temperature
glassy amorphous order would grow gradually from the liquid, and the relaxational dynamics approaching
the glass transition would be very different from the one of classical supercooled liquids.
\item The second possibility is that the glass transition line continues to be discontinuous down to $T=0$. This 
would imply the existence of a quantum discontinuous glass transition as a function of density. This has indeed been found
in~\cite{BCZ08}, based on a variational Jastrow description of the ground state of an interacting quantum particle system.
Still some conceptual problems remain: indeed, all mean field models that could be solved exactly and that display
such a transition in the classical limit, show that the latter becomes first order 
at $T=0$~\cite{MF1,MF2,MF3,MF4,MF5}.  
The model of~\cite{BCZ08} displays such a transition, but its ground state wavefunction
has only short-ranged two-body correlations, and therefore can be mapped onto a classical liquid; this is likely not
to be the case for more generic models. A better understanding of perturbation theory around Jastrow wavefunctions,
which was only attempted in \cite{BCZ08}, would be very helpful.
\end{itemize}

\section{Discussion}
\label{sec:discussion}

Before concluding the paper, we critically examine the assumptions on which our method was based:
we highlight its weaknesses and we discuss how to improve it.

The first important issue that has to be discussed is the coupling between replicas.
In section~\ref{sec:semi}, Eq.~(\ref{Hcoupled}) we have coupled the replicas as in Ref.~\cite{MP99},
which is appropriate in the classical case. Here we discuss whether this way of coupling
replicas remains correct in the quantum case.
In order to discuss this issue we go back to the original derivation of
Monasson~\cite{Mo95}. Since the glass is characterized by an inhomogeneous 
density profile, Monasson considered a system whose density is coupled to a static reference density
field $\s(r)$. We denote by
$\r(r;\{x(\t)\}) = \sum_i f(r-x_i(\t))$
a smoothed density profile at imaginary time $\t$, where the function $f(r)$ is normalized to 1 and
very short ranged. Then in the path integral representation we get
\beq
Z[\s] = \int \DD \{x\} e^{-S_0[\{x\}] - \ee \int dr d\t [\s(r) - \r(r;\{x(\t)\})]^2 } \ ,
\eeq
where $S_0[\{x\}]$ is the unperturbed action and the trajectories
$x_i$ are assumed to be periodic in $\t \to \t + \b$; a summation over the permutations due
to Bose statistics is implicit. 
Following \cite{Mo95}, we consider the partition function of the field $\s$ with ``Hamiltonian''
$F[\s] = -T \log Z[\s]$ at inverse temperature $\b m$, given by 
$Z_m= \int \DD \s(r) \exp\big[ -\b m F[\s]\big] = \int \DD \s(r) Z[\s]^m$. Integrating over $\s(r)$
we get:
\beq\label{Zdynamicfield}
Z_m  = \int \DD \{x^a\} e^{-\sum_a S_0[\{x^a\}]
- \frac{\ee}{2\b m} \int d\t d\t' dr \sum_{ab} [\r(r;\{x^a(\t)\}) - \r(r;\{x^b(\t')\})]^2 } \ .
\eeq
Defining $F(r-r') \equiv \int dz f(r-z) f(z-r')$ (note that $F(r)$ is still a normalized
short-range function), 
the coupling term has the form
\beq\label{dab}
\begin{split}
d_{ab}&(\t,\t') = \int dr [\r(r;\{x^a(\t)\}) - \r(r;\{x^b(\t')\})]^2 \\
&= \sum_{ij} [ F(x_i^a(\t)-x_j^a(\t))+ F(x_i^b(\t')-x_j^b(\t')) - 2 F(x_i^a(\t)-x_j^b(\t')) ]
\ ,
\end{split}
\eeq
This coupling makes the configuration $x_i^a(\t)$ similar to
$x_j^b(\t')$ or to a permutation of it.
Thus, in the strong coupling limit, the particles build ``molecules'' made of one atom of each
replica. 
{\it If we make the approximation of ignoring permutations of the atoms of one replica belonging
to different molecules}, or in other word if we only allow permutation of entire molecules
(molecules are Bosons but individual
atoms cannot be exchanged), then we can re-label the atoms in such a way that molecule 1 is built by
$x_1^a$, and so on. In (\ref{dab}), due to the short range of $F(r)$ we can
neglect all terms that couple atoms with $i\neq j$ and we get
\beq\nonumber
\begin{split}
d_{ab}(\t,\t') \sim 2 \sum_{i} [ F(0) - F(x_i^a(\t)-x_i^b(\t')) ] 
 \sim -\sum_{i} F''(0) (x_i^a(\t)-x_i^b(\t'))^2
\ ,
\end{split}
\eeq
where we used that $F(r)$ is short-ranged and the coupling is large, so that 
atoms in a molecule will vibrate around the
bottom of the potential well defined by $F(r)$, and we can assume that $x_i^a(\t)-x_i^b(\t')$ is small;
the result is that the coupling in a molecule can be assumed to be harmonic (note that $F''(0) < 0$).
Inserting this result in the path integral~(\ref{Zdynamicfield}), we almost recover the
path integral of the molecular system described by the Hamiltonian (\ref{Hcoupled}), except for the fact 
that replicas are coupled at different times. We therefore have to make the further approximation of
{\it ignoring the coupling of replicas at different imaginary times}.

The above derivation of our starting point, Eq.~(\ref{Hcoupled}), is then based on two crucial approximations:
\begin{itemize}
\item {\it We neglected exchange of
atoms between different molecules}. This is 
incorrect when exchange is important. But taking into account these processes, the molecules loose
their identity. A coupling term such as the one in (\ref{Hcoupled}) cannot be written since it
explicitly breaks permutation symmetry of each replica. Therefore one should write a coupling
that is manifestly invariant under permutation of the atoms of each replica, independently of
the other replicas. This can be done, but solving the Hamiltonian looks a much more difficult task. 
\item {\it We neglected the non-local coupling of replicas in imaginary time}.
It is possible than since the limit $\ee \to 0$ must be taken in the end, 
the precise form of the replica coupling is not relevant.
It would be nice to perform the calculation including the non-local coupling, but this is technically
more involved since a Hamiltonian 
formulation is not possible in this case. One has to work directly with path integrals. We attempted
to perform such a computation but did not succeed.
\end{itemize}
We note that although these problems seem very serious, they should not matter in the semiclassical
regime where exchange can be neglected, and quantum fluctuations are small, so that the path integral
is over a short time and coupling the replicas at the same time or at different times should not change
the result a lot. Therefore, we hope that Eq.~(\ref{Hcoupled}) is justified as the starting point of
our semiclassical calculation.

Another major source of concern is that a kinetic energy contribution to the equilibrium complexity
appears in Eq.~(\ref{SiT}). This contribution obviously vanishes in the classical case, and luckily enough
it still vanishes at first order in the semiclassical expansion. However, it is definitely
non-zero beyond this order. This is problematic since it would imply that 
the kinetic energy of the liquid and the glassy states visited at $m=1$ are not the same,
which is instead a basic premise of the replica approach to glasses (all correlation
functions should actually be the same). The issue is 
not the replica approach since the original Monasson's derivation 
can be formally generalized to the quantum case by using the density field $\rho(r;\{x(\tau)\})$
without encountering any difficulty, as we discussed above. The problem instead relies in the hypothesis behind 
the cage expansion, in particular the harmonic approximation. It is not clear to us 
how to find a solution. Possibly, a more refined description
of the liquid and a more accurate expansion in $\epsilon$ should 
at least reduce the problem quantitatively.

\section{Conclusions}
\label{sec:conclusions}

We presented a first attempt to build a replica theory of quantum glasses generalizing the recipe of~\cite{MP99} to the quantum case,
and performing a semiclassical expansion that seems to give meaningful results at least
for high density. In particular, we estimated the glass transition line and found that,
as already reported in~\cite{QMCT,FSZ11,Za11,FSZ10}, quantum fluctuations promote glass formation. 
We critically discussed some issues related to our approach, in particular the role of 
exchange and a technical problem related to the kinetic energy. Although a full theory
of quantum glasses must include these effects to provide meaningful results, they 
are not expected to be important in the semiclassical regime we focused on. \\
Based on the combination of the present study and previous ones~\cite{QMCT,FSZ11},
we conjectured a phase diagram of dense metastable Helium 4 which is reported in Fig.~\ref{fig:PDHEcon} (left panel).
In this case, it might well be that
the glass ceases to exist in the strongly quantum region, since it undergoes a first order transition
to the superfluid; then, the present treatment could be appropriate to describe the glass,
and the problem would be then to estimate the free energy of the dense metastable liquid in order
to look for a first order transition.
We discussed in Fig.~\ref{fig:PDHEcon2} (right panel) another possible scenario, in which the glass phase becomes
a superglass and a zero-temperature glass transition exist. The nature of this quantum 
glass transition remains unclear.

There are several ways to go beyond our treatment.
\begin{itemize}
\item One could try to use influence functional methods, which were used in~\cite{QMCT} to compute the thermodynamical
properties of the liquid. Although these methods do not include exchange, they should allow for a more accurate
description of the glass.
\item Another possibility would be to use finite-temperature variational methods~\cite{SRKC86} in order to extend
the computation of \cite{BCZ08} to finite temperature. This allows one to include exchange, but not to describe the normal phase correctly.
\item A careful investigation of the perturbation theory around the special model of \cite{BCZ08} should help to understand
if a quantum discontinuous glass transition indeed takes place.
\end{itemize}
However, since the physical picture is very far from being settled, before pushing forward the theory 
it would be extremely interesting to perform 
Quantum Monte Carlo simulations or
experiments (along the lines of \cite{supercooled}) in the dense
metastable liquid region, to investigate the behavior of the $\l$-line at high density.

\begin{acknowledgements}
We warmly thank S.Balibar, B.Clark, L.Foini, D.Reichman, G.Semerjian and M.Tarzia for many useful discussions.
\end{acknowledgements}

\end{document}